\documentclass[3p,times,twocolumn]{elsarticle}
 \biboptions{comma,sort&compress}
 
\usepackage{graphicx}
\usepackage{here}
\usepackage{ecrc}

\usepackage{graphicx,epsfig}
\usepackage{simplewick}
\usepackage{slashed}
\usepackage{color}
\usepackage{setspace}
\usepackage{hyperref}
\usepackage{empheq}

\volume{00}

\firstpage{1}

\journalname{Nuclear and Particle Physics Proceedings}

\runauth{}

\jid{nppp}

\jnltitlelogo{Nuclear and Particle Physics Proceedings}

\usepackage{amssymb}
\usepackage[figuresright]{rotating}

\def\al{\alpha}
\def\nn{\nonumber}
\def\lQ{\Lambda_{\rm QCD}}
\def\be{\begin{equation}}
\def\ee{\end{equation}}
\def\bea{\begin{eqnarray}}
\def\eea{\end{eqnarray}}
\newcommand{\eq}[1]{Eq.~\eqref{#1}}
\newcommand{\eqs}[2]{Eqs.~\eqref{#1} and \eqref{#2}}

\newcommand{\m}{{\overline m}}
\newcommand{\MS}{\overline{\rm MS}}

\begin{document}

\begin{frontmatter}

%%
%%%%%%%%%%%%%%%%%%%%%%%%%%%%%%%%%%%%%%%%%%%%%%%%%
\title{ Hyperasymptotic approximation to the operator product expansion}
 % \corref{cor0}}
% \cortext[cor0]{Talk given at 18th International Conference in Quantum Chromodynamics (QCD 15,  30th anniversary),  29 june - 3 july 2015, Montpellier - FR}
 \author[label1]{Cesar Ayala}
%  \cortext[cor0]{FAPESP CNPq-Brasil PhD student fellow.}
%\ead{rma@if.usp.br}
\address[label1]{Department of Physics, Universidad T{\'e}cnica Federico
Santa Mar{\'\i}a (UTFSM),  
Casilla 110-V, 
Valpara{\'\i}so, Chile}
 \author[label2]{Xabier Lobregat}
 % \cortext[cor1]{PhD student.}
%\ead{fanfenos@yahoo.fr}
\address[label2]{Grup de F\'\i sica Te\`orica, Dept. F\'\i sica and IFAE-BIST, Universitat Aut\`onoma de Barcelona,
E-08193 Bellaterra (Barcelona), Spain}
 \author[label2]{Antonio Pineda}
  % \fntext[fn1]{Speaker, Corresponding author.}
  %  \ead{snarison@yahoo.fr}

\pagestyle{myheadings}
\markright{ }
\begin{abstract}
These proceedings review recent work on hyperasymptotic constructions to the operator product expansion. Quantities we consider are the static potential and the pole mass. 
\end{abstract}
% \begin{document}
\begin{keyword}  
%% keywords here, in the form: keyword \sep keyword

%% MSC codes here, in the form: \MSC code \sep code
%% or \MSC[2008] code \sep code (2000 is the default)

\end{keyword}

\end{frontmatter}
%%%%%%%%%%%%
%\vspace*{-1.5cm}
\section{Introduction}

Non-perturbative (NP) effects are dominant for QCD phenomena with characteristic energy of 
${\cal O}(\lQ)$. Consequently, the absence of analytic tools for dealing with NP effects in QCD makes impossible to produce quantitative semi-analytic predictions in terms of $\lQ$ and renormalized quark masses for most low energy observables.

On the other hand, there are observables for which their perturbative expansions in powers of $\alpha$ are reasonable approximations. 
This typically happens when there is a large scale, generically referred as $Q$ ($\gg \lQ$), in the process. In principle, 
it is then possible to perform perturbative calculations up to any
finite order in $\alpha$. Nevertheless, such perturbative expansions are expected to be asymptotic and divergent. 
Such divergent behavior is not arbitrary. Besides the perturbative series in powers of $\alpha$, one also expects the observable to depend on,
non-analytic, NP, functions of order $e^{-A\frac{2\pi}{\beta_0\al(Q)}} \sim (\lQ/Q)^A$. These NP effects and the perturbative series in powers of $\alpha$ are not 
independent of each other. Indeed the former determines the late-term behavior of the later. Leaving aside instantons, that we will neglect in what follows (as they yield smaller NP corrections than those we consider in this paper), such relation 
can be quantified using the operator product expansion (OPE) of the observable for large $Q$. The allowed operators determine the allowed corrections in powers of $\lQ$ (up to logarithms), and, therefore, the large order behavior of the perturbative expansion, since the latter can be related with singularities in the Borel plane (located in the positive real axis), which mix with the NP corrections. To these singularities (and the associated asymptotic perturbative expansion) we generically refer to as infrared renormalons \cite{tHooft:1977xjm}. 

On a more general scenario one can consider more than one large scale: $Q_1 \gg Q_2 \gg \lQ$. Then the use of the OPE and the factorization between the different scales makes the perturbative expansions associated with each scale to be asymptotic. In some cases one has renormalon singularities associated with the scales $Q_1$ and $Q_2$ that cancel among themselves. This is indeed the case for the leading renormalon singularity of the pole mass and the static potential, as first found in \cite{Pineda:1998id}, and later in \cite{Hoang:1998nz,Beneke:1998rk}. We name these renormalon singularities spurious. 

So, in general, we want to:
\begin{enumerate}
\item
Predict observables with $e^{-A\frac{2 \pi }{\beta_0 \alpha(Q)}}$ precision.
\item
Avoid spurious renormalon problems.
\end{enumerate}
In this paper we focus on 1), though our results will be relevant for 2) too. 

Besides its intrinsic theoretical interest, the asymptotic behavior of perturbative expansions in QCD is starting to be seen in a series of observables, in particular, in heavy quark physics. In this case, in order to handle the renormalon problem associated with the pole mass, different threshold masses have been introduced \cite{Bigi:1994em,Beneke:1998rk,Pineda:2001zq,Lee:2003hh,Hoang:2009yr,Brambilla:2017hcq}. Some of these threshold masses introduce (explicitly or implicitly) a scale $\nu_f$ that acts as an infrared cutoff. Such infrared cutoff kills the renormalon behavior of the perturbative series producing a convergent perturbative series and introducing a linear power-like dependence in $\nu_f$. In practice these threshold masses work quite well. The error associated to the fact that we have this linear cutoff is typically small (see, for instance, \cite{Ayala:2014yxa,Ayala:2016sdn,Peset:2018ria}). Still, it is not optimal conceptually\footnote{In the same way that there is nothing conceptually wrong in using cutoff regularization in perturbative computations, but regularizations that kill spurious power-like divergences, like dimensional regularization, and preserve more symmetries are much more convenient.}. Other of these  threshold masses use approximate expressions for the Borel transform of the pole mass that partially incorporate the renormalon singularities in the Borel plane. The inverse of the Borel transform (which we will name Borel sum or Borel integral in the following) is then ill defined. This requires using some prescription to regulate the Borel integral. In this last case the perturbative series is typically abandoned and one directly works with the Borel integral expression. In this approach it is not quantified what is the error made by using (the unavoidably) approximated expressions for the Borel transform. 

This discussion leads us to consider an alternative method that is also often used to tame the asymptotic behavior of the perturbative series: truncating the perturbative sum at the minimal term. In mathematical literature, such approximation is often named the {\it superasymptotic} approximation of the original function (see \cite{BerryandHowls}), which is a name we will also use in the following. This procedure has long since been used (see \cite{Dingle},
 or \cite{LeGuillou:1990nq}, for references), mainly in the context of solutions to one-dimensional differential equations (see \cite{Boyd99}). Nevertheless, in that context, renormalons do not show up, nor it does the issue of scheme/scale dependence. 
 
In the context of four dimensional quantum gauge field theories, truncation of the perturbative sum in different formulations or using approximated expressions for the Borel integrals has also been considered since the early days of OPE/renormalon analyses to determine observables with NP accuracy (see for instance, \cite{DiGiacomo:1981lcx,LeGuillou:1990nq,Mueller:1993pa,Neubert:1994vb,Altarelli:1994vz,Ball:1995ni,Martinelli:1996pk,Broadhurst:2000yc}). However, it was not possible to make quantitative analyses beyond the large-$\beta_0$ approximation, since the existing perturbative series were only known to low orders. More recently, perturbative expansions have been obtained to high enough orders for some observables in the lattice scheme \cite{Bauer:2011ws,Bali:2013pla,Bali:2013qla,Bali:2014fea}. This permitted to quantitatively use perturbative sums truncated at the minimal term and successfully determine the gluon condensate and $\bar \Lambda$ in the quenched approximation \cite{Bali:2014sja}. This success motivates us to try to improve this approach, and to revisit with it observables already computed in the $\MS$ scheme, even if only few coefficients are known, since in the $\MS$ scheme (and in particular in heavy quark physics) renormalon dominance shows up at relatively low orders. 
 
Whereas, by construction, the superasymptotic approximation does not explicitly introduce the factorization scale $\nu_f$, the dependence on the renormalization scale $\nu$ remains to be assessed. Therefore, to push this method forward we need to get a quantitative understanding of the error on the truncation of the sum and of its remaining scheme and scale dependence. Similarly, 
the NP power corrections are potentially dependent on how the divergent perturbative series is regulated and on the renormalization scheme/scale used to define the strong coupling: $\al_X(\mu)$. A major point of this paper is to be able to control (in an analytic way) the dependence of the power corrections in this generalized scheme dependence. 
We will only then be able to add NP power corrections to the pertubative series in a systematic way, since the mixing between the perturbative series and the leading NP terms (or between the perturbative series associated to the scales $Q_1$ and $Q_2$) makes impossible to determine them independently. An unambiguous definition of the NP power corrections requires defining the perturbative series with power accuracy. Such combined expansion of perturbative series and NP terms will be called {\it hyperasymptotic} expansion as in \cite{BerryandHowls}. Organizing the computation in this way allows us to precisely state the parametric accuracy of the result at each step.

In \cite{HyperI,HyperMass} we developed such program and studied observables characterized by having a large scale $Q \gg \lQ$, and for which the operator product expansion (OPE) is believed to be a good approximation. We computed them within an hyperasymptotic expansion. More specifically, the perturbative part of the OPE was summed up using the principal value (PV) prescription: $S_{\rm PV}$, which is renormalization scale and scheme independent, as discussed in \cite{HyperI}. The difference between $S_{\rm PV}$ and the full NP result is assumed to exactly scale as the intrinsic NP terms of the OPE. In general terms:
\bea
\label{Obtruncated}
&&
{\rm Observable}(\frac{Q}{\lQ})
=
S_{\rm PV}(\al(Q))
\\
\nn
&&
+K_X^{\rm (PV)}\al_X^{\gamma}(Q)\frac{\Lambda_X^{d}}{Q^d}\left(1+{\cal O}(\al_X(Q))\right)
+{\cal O}(\frac{\Lambda_X^{d'}}{Q^{d'}})
\,,
\eea 
where the last term refers to genuine higher order terms in the OPE ($d' > d>0$).
Then, since $S_{\rm PV}$ can not be computed exactly, we obtain it approximately along an hyperasymptotic expansion (a combination of (truncated) perturbative sums and of NP corrections). This is possible if enough terms of the perturbative expansion are known, and if the divergent structure of the leading renormalons of the observable is also known. This allows us to have a clear (parametric) control on the error of the computation. Two alternative methods were considered in \cite{HyperI,HyperMass} depending on how the truncation of the perturbative sum is made. 
\begin{enumerate}
\item[{1)}]
$N$ and $\mu \sim Q$ large but finite:
\be
\label{eq:NP}
N=N_P \equiv |d|\frac{ 2\pi}{\beta_0\alpha_X(\mu)}\big(1-c\,\alpha_X(\mu)\big)
\,,
\ee
\item[{2)}]
$N \rightarrow \infty$ and $\mu \rightarrow \infty$ in a correlated way. 
\end{enumerate}
Here we only review results obtained with method 1).  Note that $c$ can partially simulate changes on the scale or scheme 
of $\alpha_X$. $d$ is the dimension associated to a given renormalon. Note that  $d$ can be positive (infrared renormalons) or negative (ultraviolet renormalons). In \cite{HyperI} only positive $d$'s were considered. 

We now give the general expression for $S_{\rm PV}(Q)$:
\be
\label{SPV}
S_{\rm PV}(Q)=S_P+\sum_{\{|d|\}}S_{|d|}+\sum_{\{d>0\}}\Omega_d+\sum_{\{d<0\}}\Omega_d
\,,
\ee
where
\be
S_P\equiv \sum_{n=0}^{N_P(|d_{min}|)}p_n\al^{n+1}(\mu)\equiv S_{|d|=0}
\,,
\ee
and ($|d| >0$)
\be
S_{|d|}=\sum_{n=N_P(|d|)+1}^{N_P(|d'|)}(p_n-p_n^{(as)})\al^{n+1}(\mu)
\,,
\ee
where the asymptotic behavior associated to renormalons with dimensions $\leq |d|$ is included in $p_n^{(as)}$, and $d'$ is the dimension of the closest renormalon to the origin in the Borel plane fulfilling that $|d'| > |d|$. 
$\Omega_d$ is the terminant \cite{Dingle} of the perturbative series associated to the singularity located at $u\equiv \frac{\beta_0 t}{4\pi}=\frac{d}{2}$ in the Borel plane. For the case of infrared renormalons ($d>0$) the general analytic expression of $\Omega_d$ can be found in \cite{HyperI} and for the ultraviolet renormalons ($d<0$) the general analytic expression can be found in \cite{HyperMass}. 

$S_{\rm PV}$ will be computed truncating the hyperasymptotic expansion in a systematic way. This means truncating \eq{SPV} as follows (note that we always define $D$ to be positive): 
\bea
\label{SPVDN}
S^{(D,N)}_{\rm PV}(Q)&=&\sum_{\{|d|\}}S_{|d|<D}+\sum_{\{|d|\leq D\}}\Omega_d
\\
\nn
&&
+
\sum_{n=N_P(D)+1}^{N_P(D)+N}(p_n-p_n^{(as)})\al^{n+1}(\mu)
\,.
\eea
For each value of the couple $(D,N)$, we can state the parametric accuracy of $S^{(D,N)}_{\rm PV}(Q)$. For instance for $S^{(0,N_P)}$ the error would be (up to a numerical  and a $\sqrt{\al_X}$ factor)
\be
\label{error0NP}
\delta S^{(0,N_P)} \sim {\cal O}\left(e^{-|d_{min}|\frac{2\pi}{\beta_0\alpha_X(Q)}}\right)
\,,
\ee
and for $S^{(|d_{min}|,0)}$  (up to a numerical  and a possible $\al^{3/2}_X$ factor):
\be
\label{errordmin}
\delta S^{(|d_{min}|,0)} \sim {\cal O}\left(e^{-|d_{min}|\frac{2\pi}{\beta_0\alpha_X(Q)}\left(1+\ln(|d/d_{min}|\right)}\right)
\,,
\ee
where $d$ is the location of the next renormalon closest to the origin. This corresponds to the first term in the hyperasymptotic approximation. The expression for the error in the general case $S^{(D,N)}_{\rm PV}(Q)$ reads ($N \not=N_P$ but large)
\be
\label{errorgeneral}
\delta S^{(D,N)} \sim {\cal O}\left(e^{-D\frac{2\pi}{\beta_0\alpha_X(Q)}\left(1+\ln(|d/D|\right)}\al_X^N\right)
\,,
\ee
where $d$ is the location of the next renormalon closest to the origin after $D$.

We next discuss the static potential and the pole mass in the large $\beta_0$ approximation where we can quantitatively test these ideas.

\section{The static potential in the large $\beta_0$ approximation}
The QCD static potential is written in terms of its Fourier transform as 
\be
V(r)=-\frac{2C_F}{\pi}\int_0^\infty dq \frac{\sin qr}{qr}\alpha_v(q)\ .
\label{Vr}
\ee
This equation defines $\alpha_v(q)$ in the V-scheme. In the large-$\beta_0$ approximation, we know the behavior of $\alpha_v(q)$ as a series in powers of $\alpha_X\equiv\alpha_X(\mu)$
\be
\alpha_v(q)=\alpha_X\sum_{n=0}^\infty L^n=\alpha_X\frac{1}{1-L}\ ,
\label{alsV}
\ee
where $L=\frac{\beta_0\alpha_X}{2\pi}\ln(\frac{\mu e^{-c_X/2}}{q})$. If $X=\MS$ then $c_{\MS}=-5/3$ (in the large $\beta_0$ approximation). If $X=V$ then $c_V=0$.  If $X={\rm latt}$, we take the $n_f=0$ number for a Wilson action of $d_1=5.88359$ \cite{Hasenfratz:1980kn} and use $c_{\rm latt}=-2(\frac{5}{6}+\frac{2\pi d_1}{\beta_0})$, as we only use this scheme for checking the consistency between the results obtained with different schemes. Note that this yields two values of $c_{\rm latt}$ if we introduce the $n_f$ dependence of $\beta_0$: $c_{\rm latt}(n_f=0)=-8.38807$ and $c_{\rm latt}(n_f=0)=-9.88171$. This is the way we have implemented the large $\beta_0$ approximation in the lattice scheme in \cite{HyperI} and \cite{HyperMass}. We also define $\tilde \Lambda 
= \Lambda_X e^{-c_X/2}$ and $\rho=\tilde \Lambda r$. 
Note that $\tilde \Lambda$ is scheme independent. 
 
\eq{Vr} is ill defined but not its Borel transform. It reads \cite{Aglietti:1995tg}
\be
B[V](t(u))=B(t(u))=\frac{-C_F}{\pi^{1/2}}\frac{1}{r} e^{-c_Xu}\bigg(\frac{\mu^2r^2}{4}\bigg)^u\frac{\Gamma(1/2-u)}{\Gamma(1+u)}
\,,
\ee
which is a meromorphic function in the $u$ complex plane. 

We then define (where the single poles of the Borel transform are regulated using the PV prescription)
\be
V_{\rm PV}(r)=\int_{0,\rm PV}^{\infty} dt e^{-t/\alpha(\mu)}B[V](t(u))
\,.
\ee

We will perform computations with $n_f=0$ and $n_f=3$. In the first case we will work in lattice units (aiming to compare with quenched lattice simulations) and use $\Lambda_{\MS}(n_f=0)=0.602 r_0^{-1} \approx 238$ MeV \cite{Capitani:1998mq}. 
In the large $\beta_0$ approximation (with $n_f=0$), this yields $\al(M_{\tau}) \approx 0.29$.  
In the second case we take $\Lambda_{\MS}(n_f=3)=174$ MeV. This last number we fix such that it gives a reasonable value at the $\tau$ mass in the large $\beta_0$ approximation: $\al(M_{\tau}) \approx 0.3$ (see for instance \cite{Boito:2018yvl}). We then confront $V_{\rm PV}$ with the results obtained truncating the hyperasymptotic expansion. 

We define
\be
V_P\equiv \sum_{n=0}^{N_P} V_n \al^{n+1}
\,.
\ee
Applying \eq{SPVDN} to the static potential in the large $\beta_0$ approximation, the relation between $V_{\rm PV}$ and $V_{P}$ reads
\be
\label{Vb0PV1}
V_{\rm PV}=V_{P}+\frac{1}{r}\Omega_V+\sum_{n=N_P+1}^{3N_P} (V_n-V_n^{(\rm as)}) \al^{n+1}
+\frac{1}{r}\Omega_V'+o(\lQ^3 r^2)
\,,
\ee
where $\Omega_V$ reads for this case
\bea
\label{eq:OmegaV}
\Omega_V&=&\sqrt{\alpha_X(\mu)}K_X^{(P)} r \, \mu e^{-\frac{2\pi}{\beta_0 \alpha_X(\mu)}}
\\
\nn
&&
\hspace{-0.7cm}
\times
\bigg(1+\bar K_{X,1}^{(P)}\alpha_X(\mu)+\bar K_{X,2}^{(P)}\alpha_X^2(\mu)+\mathcal{O}\left(\alpha_X^3(\mu)\right)\bigg)
\,,\eea
and $K_X^{(P)}$ and $\bar K_{X,i}^{(P)}$ read
\bea
K_X^{(P)}&=&- 2\pi  Z^X_V 
\beta_0^{-1/2}\bigg[-\eta_c+\frac{1}{3}\bigg]
\,,
\\
\bar K_{X,1}^{(P)}&=&\frac{\beta_0/(\pi)}{-\eta_c+\frac{1}{3}}\bigg[
-\frac{1}{12}\eta_c^3+\frac{1}{24}\eta_c-\frac{1}{1080}\bigg]
\,,
\\
\bar K_{X,2}^{(P)}&=&\frac{\beta_0^2/\pi^2}{-\eta_c+\frac{1}{3}}
\bigg[
-\frac{1}{160}\eta_c^5
\\
\nn
&&
\hspace{-0.7cm}
-\frac{1}{96}\eta_c^4+\frac{1}{144}\eta_c^3
+\frac{1}{96}\eta_c^2-\frac{1}{640}\eta_c-\frac{25}{24192}\bigg]
\,,
\eea
where 
\be
Z_V^{X}=-2 \frac{C_F}{\pi}e^{-\frac{c_X}{2}} \quad {\rm and} \quad \eta_c=\frac{2\pi c}{\beta_0}-1
\,.
\ee
Note that in the large $\beta_0$ we identically have $\Lambda_X=\mu e^{-2\pi/(\beta_0 \alpha_X(\mu))}$.  A similar expression applies to $\Omega'_V \sim \sqrt{\al_X(\mu)}(r\lQ)^3$. For the numerical discussion below we have used the numerical evaluation of $\Omega_V$, but using \eq{eq:OmegaV} yields the same results.

\begin{center}
\begin{figure}
\includegraphics[width=0.5\textwidth]{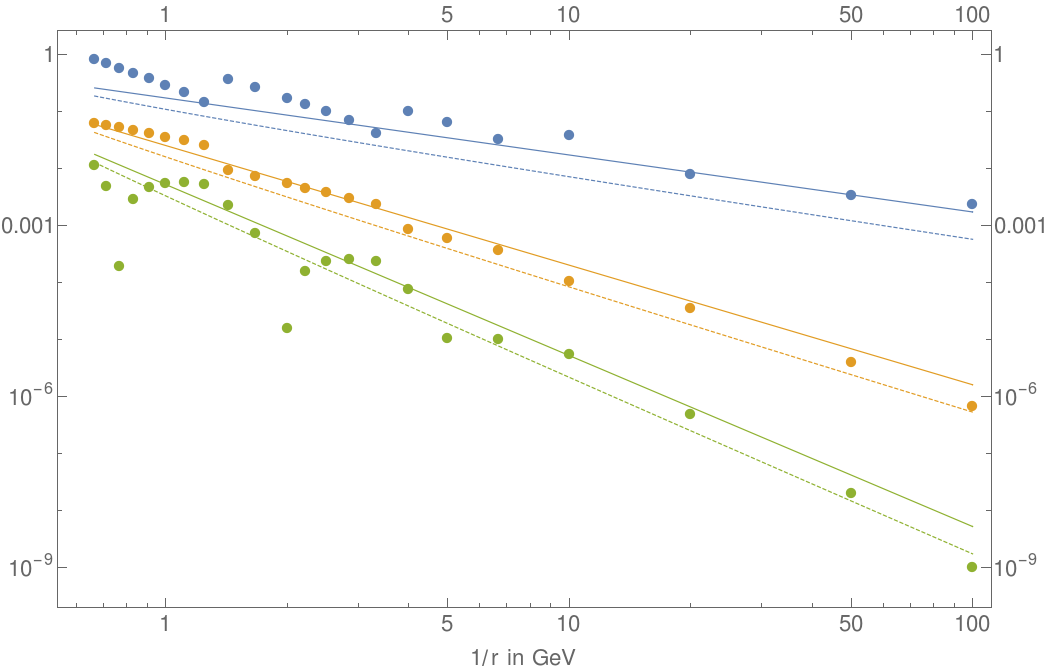}
\caption{ Blue points are $|r(V_{\rm PV}-V_P)|$. Orange points are $|r(V_{\rm PV}-V_P)-\Omega_V|$, Green 
$|r(V_{\rm PV}-V_P)-\Omega_V-r\sum_{n=N_P+1}^{3N_P} (V_n-V_n^{(\rm as)}) \al^{n+1}|$. They are plotted as functions of $1/r$ in logarithmic scale (which is equivalent to plotting them in terms of $1/\al$). The continuous lines are $e^{-\frac{2\pi}{\beta_0\al}}$ (blue), $e^{-(1+\ln 3)\frac{2\pi}{\beta_0\al}}$ (orange),   $e^{-3\frac{2\pi}{\beta_0\al}}$ (green). The dashed lines are the same functions multiplied by $\sqrt{\al}$. In the notation $(D,N)$ of \eq{SPVDN} they correspond to $(0,N_P)$, $(1,0)$ and $(1,2N_P)$ respectively. The computation has been done with $n_f=3$, in the $\MS$ scheme, and taking the smallest positive value possible of $c$ that yields integer values for $N_P$.}
\label{Fig:Fig7BoydLargeb0MS}
\end{figure}
\end{center}

\begin{center}
\begin{figure}
\includegraphics[width=0.5\textwidth]{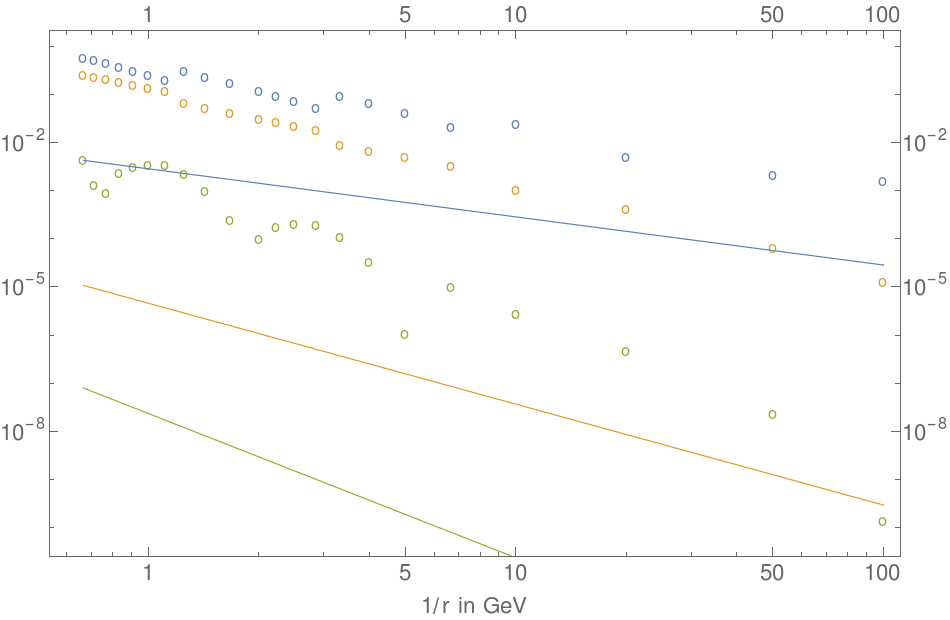}
\caption{Blue points are $|r(V_{\rm PV}-V_P)|$. Orange points are $|r(V_{\rm PV}-V_P)-\Omega_V|$, Green 
$|r(V_{\rm PV}-V_P)-\Omega_V-r\sum_{n=N_P+1}^{3N_P} (V_n-V_n^{(\rm as)}) \al^{n+1}|$. They are plotted as functions of $1/r$. The continuous lines are $e^{-\frac{2\pi}{\beta_0\al}}$ (blue), $e^{-(1+\ln 3)\frac{2\pi}{\beta_0\al}}$ (orange), $e^{-3\frac{2\pi}{\beta_0\al}}$ (green). In the notation $(D,N)$ of \eq{SPVDN} they correspond to $(0,N_P)$, $(1,0)$ and $(1,2N_P)$ respectively. The computation has been done with $n_f=3$, in the lattice scheme, and taking the smallest positive value possible of $c$ that yields integer values for $N_P$.}
\label{Fig:Fig7BoydLargeb0Latt}
\end{figure}
\end{center}

\begin{center}
\begin{figure}
\includegraphics[width=0.5\textwidth]{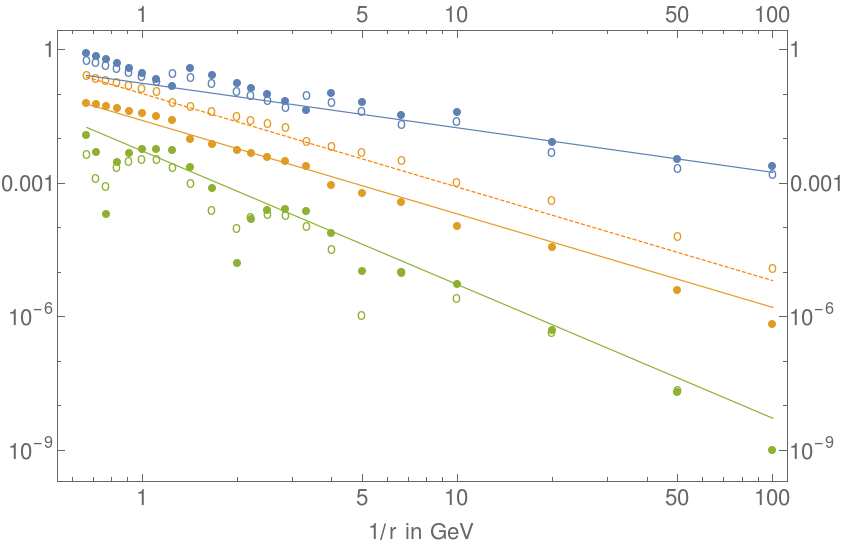}
\caption{ Blue points are $|r(V_{\rm PV}-V_P)|$. Orange points are $|r(V_{\rm PV}-V_P)-\Omega_V|$, Green 
$|r(V_{\rm PV}-V_P)-\Omega_V-r\sum_{n=N_P+1}^{3N_P} (V_n-V_n^{(\rm as)}) \al^{n+1}|$. 
They are plotted as functions of $1/r$. Full points are computed in the $\MS$ scheme and empty points in the lattice scheme. The continuous lines are $e^{-\frac{2\pi}{\beta_0\al_{\MS}}}=e^{-5/6}e^{-c_{\rm latt}/2}e^{-\frac{2\pi}{\beta_0\al_{\rm latt}}}$ (blue), $e^{-(1+\ln 3)\frac{2\pi}{\beta_0\al_{\MS}}}=e^{-(1+\ln 3)(5/6+c_{\rm latt}/2)}e^{-(1+\ln 3)\frac{2\pi}{\beta_0\al_{\rm latt}}}$ (orange),   $e^{-3\frac{2\pi}{\beta_0\al_{\MS}}}=e^{-5/2}e^{-3c_{\rm latt}/2}e^{-3\frac{2\pi}{\beta_0\al_{\rm latt}}}$ (green). In the notation $(D,N)$ of \eq{SPVDN} they correspond to $(0,N_P)$, $(1,0)$ and $(1,2N_P)$ respectively. The dashed orange line corresponds to $4\times e^{-(1+\ln 3)\frac{2\pi}{\beta_0\al_{\MS}}}$. The computation has been done with $n_f=3$ and taking the smallest positive value possible of $c$ that yields integer values for $N_P$.}
\label{Fig:Fig7BoydLargeb0MSLatt}
\end{figure}
\end{center}

We illustrate the accuracy achieved at different orders in the hyperasymptotic expansion in Fig. \ref{Fig:Fig7BoydLargeb0MS}. The difference between the exact result and the truncated hyperasymptotic expansion is consistent with the scaling of the error given above (see Eqs. (\ref{error0NP})-(\ref{errorgeneral})). The inclusion of a $\sqrt{\al}$ factor does not change the qualitative picture much (compared with the exponential behavior, $\sqrt{\al}$ changes little the slope). It should be stressed though that this comparison has been done in the $\MS$ scheme. If we do the comparison in the lattice scheme, we find that the error estimates given by Eqs. (\ref{error0NP})-(\ref{errorgeneral}) are much smaller than the real differences between the exact result and the truncated hyperasymptotic expansion. We show the results in Fig. \ref{Fig:Fig7BoydLargeb0Latt}. Some interesting considerations can be made though. In Fig. \ref{Fig:Fig7BoydLargeb0MSLatt} we display the difference between the exact result and the truncated hyperasymptotic expansion in the $\MS$ and in the lattice scheme, as full or empty points. We find that the truncations $(0,N_P)$ and $(1,2N_P)$ are quite independent of the scheme used for the perturbative expansion or, in other words, they are quite independent of the renormalization scale $\mu$ used for the strong coupling $\al$ (we remind that in the large $\beta_0$ approximation a change of scheme is equivalent to a change of renormalization scale). Therefore, in those cases, the scaling of the error estimate with $1/r$ is also qualitatively correct in the lattice case but what happens, compared with Fig. \ref{Fig:Fig7BoydLargeb0MS}, is that the overall magnitude is not of ${\cal O}(1)$ but smaller. Therefore, it would be misleading to use them for the error (the overall normalization is important). Nevertheless, one could still fix the normalization using a specific energy (i.e. for a fixed $1/r$)  and then one would have a fair estimate of the error for other energies. One could also fix the normalization using the $\MS$ scheme result where a factor of ${\cal O}(1)$ multiplying the exponential properly accounts for the error. Then one can take the lattice result and multiply it by the corresponding change of scheme, to get the proper magnitude of the error.

A slightly different discussion takes place for the $(D,N)=(1,0)$ truncation. As we see in Fig. \ref{Fig:Fig7BoydLargeb0MSLatt} there is a significant difference between the (1,0) hyperasymptotic approximation in the 
$\MS$ and the lattice scheme. We see that the $\MS$ scheme yields better results (or, equivalently, choosing a $\mu$ scale that makes $\alpha$ bigger is better) than the lattice scheme. The reason is that adding the $\Omega_V$ is a relatively small correction in the lattice scheme, which is then corrected by the longer array of perturbative terms that add up till catching up the precision obtained in the $\MS$ scheme. In any case, the $1/r$ dependence is roughly equivalent in both cases. Therefore, one  could still fix the normalization using a specific energy (i.e. $1/r$)  and then one would have a fair estimate of the error for other energies but this coefficient can not be (easily) understood as a change of scheme. 

\section{The pole mass in the large $\beta_0$ approximation}
\label{Sec:polemass}
We now consider the pole mass. Here the discussion runs parallel to the discussion for the static potential in the large $\beta_0$ approximation. Nevertheless, we do not have the same analytic control as for the static potential. Note also that now we have ultraviolet renormalons. Moreover, the pole mass has the extra complication that it is ultraviolet divergent and needs renormalization. This makes the Borel transform more complicated and we do not have the exact $\mu$ factorization one has in the static potential. We take the Borel transform from \cite{Beneke:1994sw,Ball:1995ni,Neubert:1994vb}:
\bea
&&B[m_{\rm PV}-\m](u)
=
\m \frac{C_F}{4\pi}\bigg[\bigg(\frac{\m^2}{\mu^2}\bigg)^{-u}e^{-c_{\MS}u}6(1-u)
\nn
\\
&&
\qquad
\times\frac{\Gamma(u)\Gamma(1-2u)}{\Gamma(3-u)}-\frac{3}{u}+R(u)\bigg]
\,,
\eea
where $u=\frac{\beta_0}{4\pi}t$ and
\be
R(u)=\sum_{n=1}^{\infty}\frac{1}{(n!)^2}\frac{d^n}{dz^n}G(z)\bigg|_{z=0}u^{n-1}=-\frac{5}{2}+\frac{35}{24}u+\mathcal{O}(u^2)
\,,
\ee
\be
G(u)=-\frac{1}{3}(3+2u)\frac{\Gamma(4+2u)}{\Gamma(1-u)\Gamma^2(2+u)\Gamma(3+u)}
\,.
\ee
This expression has been derived in the $\MS$ scheme. Whereas the scheme dependence of the first term can be reabsorbed in changes of $\mu$ and $c_{\MS}$ (it would then be equivalent to a change of scale), controlling the scheme dependence of $R(u)$ is more complicated. We will not care much, as $R(u)$ has to do with the high energy behavior, and should only affect 
\be
m_P\equiv \m+\sum_{n=0}^{N_P} r_n\al^{n+1}(\mu)\;;
\ee 
i.e. the finite sum. Therefore, when we change from the $\MS$ to the lattice scheme we will leave $R(u)$ unchanged. Strictly speaking then, the object we compute in the lattice scheme is not the pole mass, still it will have the same infrared behavior. The fact that we will obtain the same result after subtracting $m_P$ from $m_{\rm PV}$ in both cases will be a nice confirmation that high-energy cancellation has effectively taken place and what is left is low energy\footnote{To make an analogy, the situation is similar to determinations of the infrared behavior of the energy of an static source in perturbation theory. In \cite{Bauer:2011ws,Bali:2013pla,Bali:2013qla} two different discretizations were used for the static quark propagators. This affected the ultraviolet, but let the infrared behavior unchanged, as it was nicely seen in those simulations. See also the discussion in \cite{Hayashi:2019mlb}.}. 

The hyperasymptotic expansion of $m_{\rm PV}(\m)$ at low orders reads
\bea
\nn
m_{\rm PV}(\m)&=&m_P+\m \Omega_m+\sum_{n=N_P+1}^{2N_P}(r_n-r_n^{(as)})\al^{n+1}(\mu)
\\
\label{mPV}
&&
\hspace{-0.5cm}
+
\m\Omega_2+\m \Omega_{-2}+{\cal O}\left(e^{-2\frac{2\pi}{\beta_0\al}\left(1+\ln(3/2)\right)}\right)
\,,
\eea
where 
\begin{equation}
\label{Omegam}
\Omega_m=-\frac{1}{2}\Omega_V
\,,\end{equation}
replacing $1/r \rightarrow \m$.

In the large $\beta_0$ approximation $\Omega_2=0$ and 
\bea
\label{Omegaminustwo}
&&\Omega_{-2}=\sqrt{\alpha(\mu)}K_X^{(P)}
\frac{\Lambda^{2} \m^{2}}{\mu^{4}}
\\
\nn
&&\times
\bigg\{1+\bar{K}_{X,1}^{(P)}\alpha(\mu)+\bar{K}_{X,2}^{(P)}\alpha^2(\mu)+\mathcal{O}\left(\alpha^3(\mu)\right)\bigg\}
\,,
\eea
\be
K_X^{(P)}\equiv Z^X_{-2}(-1)^{N_p+1}\left(\frac{\beta_0}{\pi^22}\right)^{-1/2}\,,
\qquad
Z_{-2}^X=-\frac{C_F e^{c_X}}{\pi}
\,,
\ee
\bea
\bar{K}_{X,1}^{(P)}&\equiv&
\frac{\beta_0}{24\pi}(-1+3\eta_c^{2})
\,,
\\
\bar{K}_{X,2}^{(P)}&\equiv&
\frac{\beta_0^2}{4608 \pi^{2}}\bigg[13-48\eta_c
-60\eta_c^{2}
\\
\nn
&&
+48\eta_c^{3}+36\eta_c^{4}\bigg]
\,,
\eea
where now $\eta_c=\frac{2\pi c}{\beta_0}\times 2-1$.
For the numerical discussion below we have used the numerical evaluation of $\Omega_{-2}$, but using \eq{Omegaminustwo} yields the same results.

\begin{center}
\begin{figure}
\includegraphics[width=0.5\textwidth]{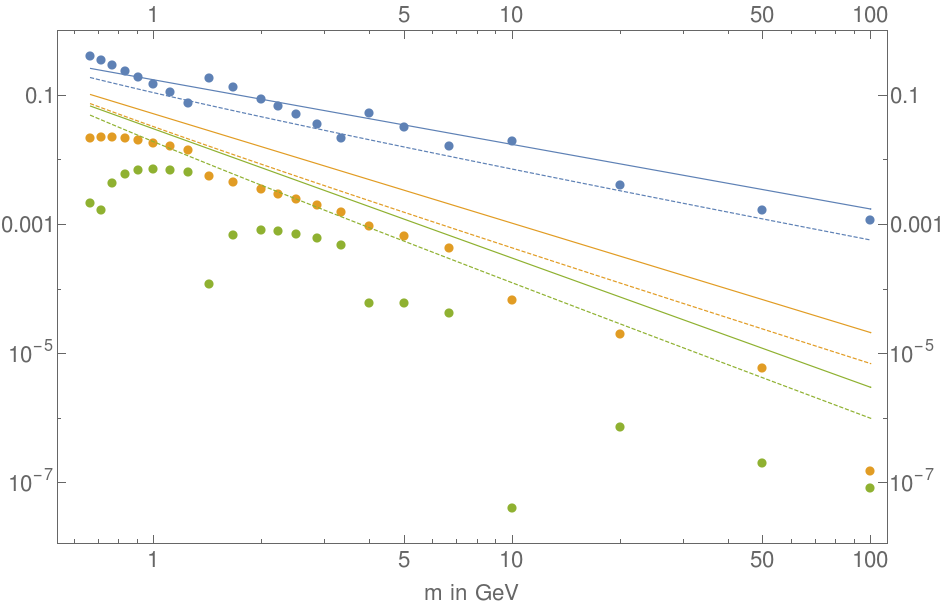}
\caption{Blue points are $|(m_{\rm PV}-m_P)/\m|$. Orange points are $|(m_{\rm PV}-m_P)/\m-\Omega_m|$, Green 
$|(m_{\rm PV}-m_P)/\m-\Omega_m-\sum_{n=N_P+1}^{2N_P} (r_n-r_n^{(\rm as)}) \al^{n+1}/\m|$. They are plotted as functions of $\m$. The continuous lines are 
$e^{-\frac{2\pi}{\beta\al}}$ (blue), $e^{-(1+\ln 2)\frac{2\pi}{\beta\al}}$ (orange),   $e^{-2\frac{2\pi}{\beta\al}}$ (green). The dashed lines are the same functions multiplied by $\sqrt{\al}$. In the notation $(D,N)$ of \eq{SPVDN} they correspond to $(0,N_P)$, $(1,0)$ and $(1,N_P)$ respectively. The computation has been done with $n_f=3$, in the $\MS$ scheme, and taking the smallest positive value possible of $c$ that yields integer values for $N_P$.}
\label{Fig:Fig7BoydLargeb0MSmass}
\end{figure}
\end{center}

\begin{center}
\begin{figure}
\includegraphics[width=0.5\textwidth]{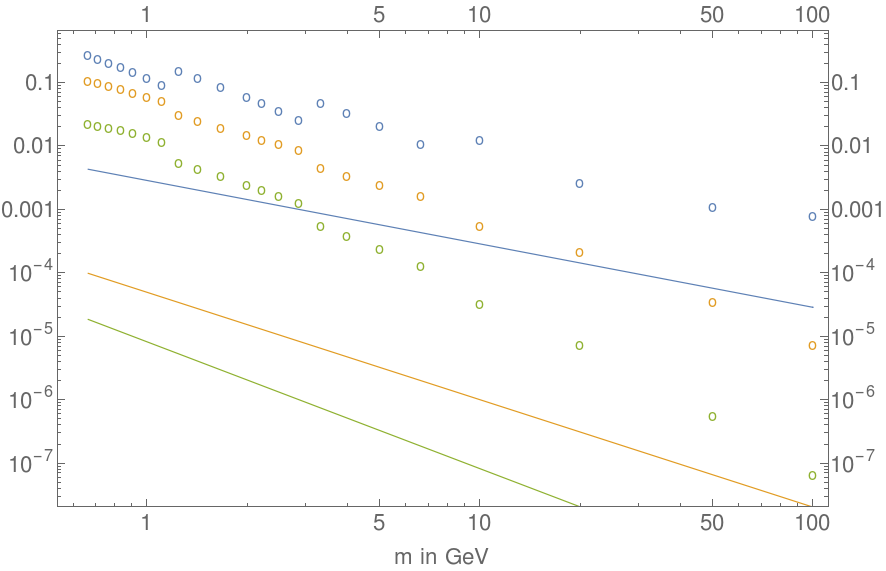}
\caption{Blue points are $|(m_{\rm PV}-m_P)/\m|$. Orange points are $|(m_{\rm PV}-m_P)/\m-\Omega_m|$, Green 
$|(m_{\rm PV}-m_P)/\m-\Omega_m-\sum_{n=N_P+1}^{2N_P} (r_n-r_n^{(\rm as)}) \al^{n+1}/\m|$. They are plotted as functions of $m$. The continuous lines are $e^{-\frac{2\pi}{\beta_0\al}}$ (blue), $e^{-(1+\ln 2)\frac{2\pi}{\beta_0\al}}$ (orange), $e^{-2\frac{2\pi}{\beta_0\al}}$ (green). In the notation $(D,N)$ of \eq{SPVDN} they correspond to $(0,N_P)$, $(1,0)$ and $(1,N_P)$ respectively. The computation has been done with $n_f=3$, in the lattice scheme, and taking the smallest positive value possible of $c$ that yields integer values for $N_P$.}
\label{Fig:Fig7BoydLargeb0LattMass}
\end{figure}
\end{center}

\begin{center}
\begin{figure}
\includegraphics[width=0.5\textwidth]{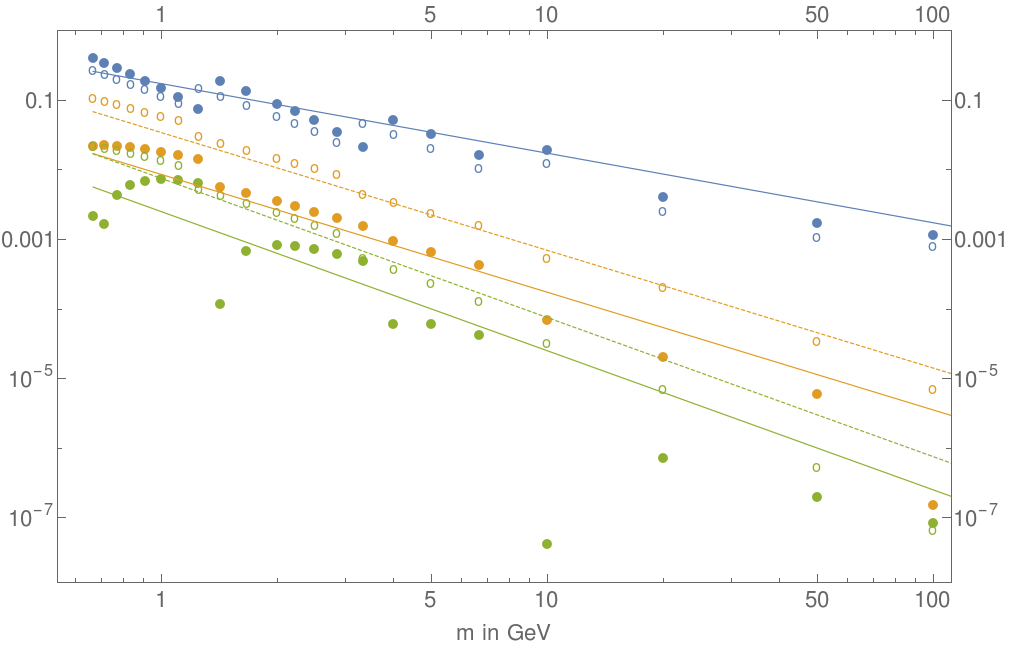}
\caption{ Blue points are $|(m_{\rm PV}-m_P)/\m|$. Orange points are $|(m_{\rm PV}-m_P)/\m-\Omega_m|$, Green 
$|(m_{\rm PV}-m_P)/\m-\Omega_m-\sum_{n=N_P+1}^{2N_P} (r_n-r_n^{(\rm as)}) \al^{n+1}/\m|$. 
They are plotted as functions of $\m$. Full points are computed in the $\MS$ scheme and empty points in the lattice scheme. The continuous lines are $e^{-\frac{2\pi}{\beta_0\al_{\MS}}}$ (blue), $\frac{1}{6}e^{-(1+\ln 2)\frac{2\pi}{\beta_0\al_{\MS}}}$ (orange),   $\frac{1}{12}e^{-2\frac{2\pi}{\beta_0\al_{\MS}}}$ (green). In the notation $(D,N)$ of \eq{SPVDN} they correspond to $(0,N_P)$, $(1,0)$ and $(1,N_P)$ respectively. The dashed orange and green lines correspond to $\frac{2}{3}\times e^{-(1+\ln 2)\frac{2\pi}{\beta_0\al_{\MS}}}$ and $\frac{1}{4}\times e^{-2\frac{2\pi}{\beta_0\al_{\MS}}}$ respectively. The computation has been done with $n_f=3$ and taking the smallest positive value possible of $c$ that yields integer values for $N_P$.}
\label{Fig:Fig7BoydLargeb0MSLattMass}
\end{figure}
\end{center}

\begin{center}
\begin{figure}[htb]
\includegraphics[width=0.5\textwidth]{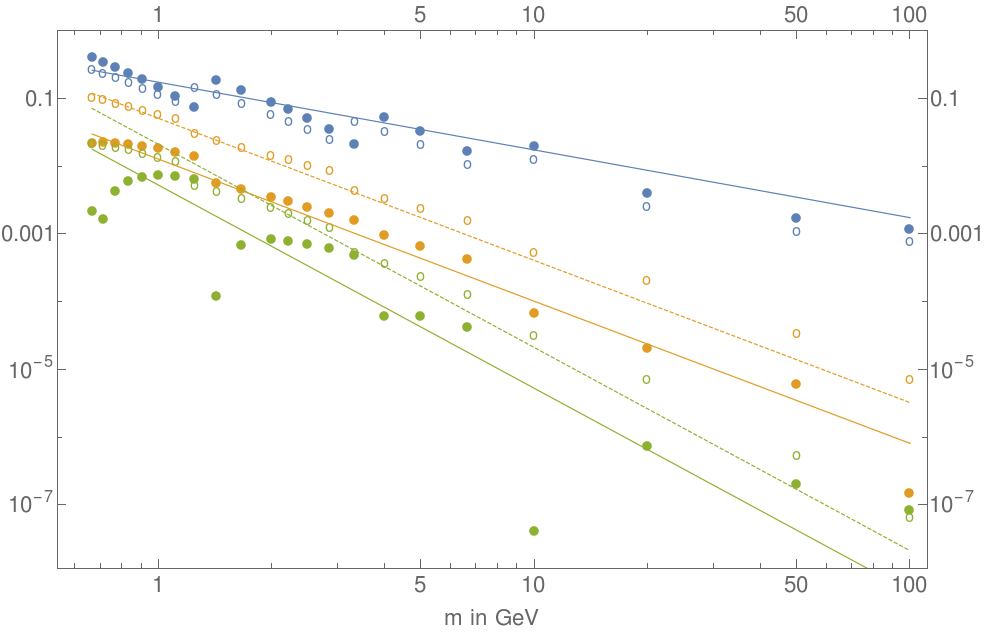}
\caption{ Blue points are $|(m_{\rm PV}-m_P)/\m|$. Orange points are $|(m_{\rm PV}-m_P)/\m-\Omega_m|$, Green 
$|(m_{\rm PV}-m_P)/\m-\Omega_m-\sum_{n=N_P+1}^{2N_P} (r_n-r_n^{(\rm as)}) \al^{n+1}/\m|$. 
They are plotted as functions of $\m$. Full points are computed in the $\MS$ scheme and empty points in the lattice scheme. The continuous lines are $e^{-\frac{2\pi}{\beta_0\al_{\MS}}}$ (blue), $\frac{1}{2}e^{-(1+\ln 3)\frac{2\pi}{\beta_0\al_{\MS}}}$ (orange),   $e^{-3\frac{2\pi}{\beta_0\al_{\MS}}}$ (green). The dashed orange and green lines correspond to $2\times e^{-(1+\ln 3)\frac{2\pi}{\beta_0\al_{\MS}}}$ and $4\times e^{-3\frac{2\pi}{\beta_0\al_{\MS}}}$ respectively. The computation has been done with $n_f=3$ and taking the smallest positive value possible of $c$ that yields integer values for $N_P$.}
\label{Fig:Fig7BoydLargeb0MSLattMass3}
\end{figure}
\end{center}

We now illustrate the accuracy achieved at different orders in the hyperasymptotic expansion in the $\MS$ scheme of the pole mass in Fig. \ref{Fig:Fig7BoydLargeb0MSmass}. There are similarities as well as differences with the previous discussion of the static potential. We find that the truncation $(D,N_P)=(0,N_P)$ yields an error consistent with \eq{error0NP}, as it also happened for the static potential. The situation for the other truncations is different. The next renormalon now is of ultraviolet origen. This renormalon is multiplied by an small number compared with the case of the infrared renormalon. This makes that the error given by \eqs{errordmin}{errorgeneral} (which take a coefficient of ${\cal O}(1)$) is too large compared with the real error of the truncation. This happens both for the truncations (1,0) and $(1,N_P)$. We next consider the same analysis but in the lattice scheme. We show the results in Fig. \ref{Fig:Fig7BoydLargeb0LattMass}. In this case we have that the real difference between the truncation of the hyperasymptotic expansion and the exact result are always larger than the estimates in Eqs. (\ref{error0NP}), (\ref{errordmin}) and (\ref{errorgeneral}). In any case one may wonder whether the overall coefficient can be fixed in all cases such that the error of the truncation can be well parameterized by the error formulas given in Eqs. (\ref{error0NP}), (\ref{errordmin}) and (\ref{errorgeneral}). We perform a first preliminary analysis. We show tentatives lines in Fig. \ref{Fig:Fig7BoydLargeb0MSLattMass}. The lines for the truncations (1,0) and $(1,N_P)$ do not accurately follow the real difference between the exact result and the truncated sum. The reason for this discrepancy should be understood better. In any case, it seems to be linked to the fact that we are dealing with ultraviolet renormalons here. The smallness of the normalization of those could produce that in practice the error is dominated by the next infrared renormalon or, being more precise, depending on the scale (and scheme) the $u=-1$ or the $u=3/2$ dominates.  The possibility that the error is dominated by the $u=3/2$ renormalon is indeed supported by the fact that the slope of the points describing the difference between the exact result and the hyperasymptotic expansion is bigger than the one predicted by the $u=-1$ renormalon. We indeed explore this possibility in Fig. \ref{Fig:Fig7BoydLargeb0MSLattMass3}. For the $\MS$ scheme this option does a decent job, whereas for the lattice it overshoots a little bit. Note, still, that if we are only interested in the error (and we fix the normalization at some fixed $m$), to assume that the $u=-1$ renormalon dominates would only imply that we are using a conservative estimate of the error (and the real error would be smaller). 

\section{Conclusions}
In these proceedings we have reviewed the application of hyperasymptotic approximations to observables where the OPE can be applied, and that suffer from renormalon ambiguities. We have particularly focused on the static potential and the pole mass. We have computed both of them in the large $\beta_0$ approximation. They are interesting to study because they allow us to see the structure of the hyperasymptotic expansion in full glory. 

For the case of the static potential, where there are only infrared renormalons, it is nicely seen (see particularly Fig. \ref{Fig:Fig7BoydLargeb0MSLatt}) that the scaling of the error in $1/r$ perfectly fits with the theoretical formulas given in Eqs. (\ref{error0NP}), (\ref{errordmin}), and (\ref{errorgeneral}). This scaling is indeed independent of the scheme. One can also see that the overall normalization of the error can not be fixed a priori (they would require of more dedicated theoretical studies). Nevertheless, what we find is that for truncations at each minimal term of the hyperasymptotic expansion, in our case $(0,N_P)$ and $(1,2N_P)$, the normalization is quite independent of the scheme. For those cases the normalization is of natural size if working in the $\MS$ scheme. Therefore, for the lattice scheme, the normalization is large (indeed it corresponds to the change of scheme from the lattice to the $\MS$ scheme). Significantly, the truncation $(1,0)$ shows a different behavior with respect the truncations $(0,N_P)$ and $(1,2N_P)$. As we already said, in both schemes the error scales with $1/r$ as expected for the $(1,0)$ truncation. Nevertheless, the normalization shows a dependence on the scheme.  In this respect we observe a better convergence working in the $\MS$ scheme than in the lattice scheme (in other words the normalization factor is smaller in the $\MS$ than in the lattice scheme). 

For the case of the pole mass, besides infared renormalons (which behave similarly to the case of the static potential), we have ultraviolet renormalons. These show a different behavior. This is due to the fact that the normalization of those is much smaller than the normalizations one has for infrared renormalons. Therefore, even if they are nominally more important than subleading infrared renormalons, they may mix in practice (or even the infrared renormalon can be more important). To be in one situation or another depends on the scale of the problem as well on the scheme used for the computation. This is discussed in more detail in the discussion at the end of Sec. \ref{Sec:polemass} and in the associated figures. 

Other observables have been considered using a hyperasymptotic approximation in \cite{HyperMass}, in particular the pole mass beyond the large $\beta_0$ limit, which has then been used to get estimates of the error of determinations of the $\m$ bottom mass from the $B$ meson mass, and the error of determinations of the $\m$ top mass if the PV of the top mass is known. Actually, tiny errors were obtained. For instance, for the top quark, the error associated to approximating the PV mass by the present knowledge of its hyperasymptotic expansion was estimated to be of order 
$\sim 30$ MeV. Besides the pole mass, though directly related to it, determinations of $\bar \Lambda$ have been considered in the lattice, as well as a study of the renormalon dependence of the static potential at a fixed $r$ (in other words of the dependence of the static potential in the lattice spacing).  

\medskip

\noindent
{\bf Acknowledgments}\\
\noindent
This work was supported in part by the Spanish FPA2017-86989-P and SEV-2016-0588 grants from the ministerio de Ciencia, Innovaci\'on y Universidades, and the 2017SGR1069 grant from the Generalitat de Catalunya; and by the Chilean FONDECYT Postdoctoral Grant No. 3170116, and by FONDECYT Regular Grant No. 1180344.

\end{document}